\documentclass{elsart}
\usepackage{epsfig,amssymb,bm}
\begin{document}

\begin{frontmatter}
\title{$\bm{\mu}$SR and NMR in f-electron non-Fermi liquid materials}
\author[UCR]{\underline{D. E. MacLaughlin},}
\author[UCR]{M. S. Rose,}
\author[UCR]{Ben-Li Young,}
\author[CSU]{O. O. Bernal,}
\author[LANL]{R. H. Heffner,}
\author[LANL]{G. D. Morris,}
\author[UCR,OU]{K. Ishida,\thanksref{KU}}
\author[KOL]{G. J. Nieuwenhuys,}
\and \author[LANL,SFU]{J. E. Sonier}
\thanks[KU]{Present address: Department of Physics, Kyoto University, Kyoto 606-8502, Japan.}

\address[UCR]{Department of Physics, University of California, \\Riverside, California 92521, U.S.A.}
\address[CSU]{Department of Physics and Astronomy, California State University,\\ Los Angeles, California 90032, U.S.A}
\address[LANL]{MS K764, Los Alamos National Laboratory, \\Los Alamos, New Mexico 87545, U.S.A.}
\address[OU]{Department of Physical Science, Graduate School of Engineering Science, \\Osaka University, Toyonaka, Osaka 560-8531, Japan}
\address[KOL]{Kamerlingh Onnes Laboratorium, Leiden University, \\2300 RA Leiden, The Netherlands}
\address[SFU]{Department of Physics, Simon Fraser University,\\ Burnaby, British Columbia, Canada V5A 1S6}

\begin{abstract}
Magnetic resonance ($\mu$SR and NMR) studies of f-electron non-Fermi-liquid (NFL) materials give clear evidence that structural disorder is a major factor in NFL behavior. Longitudinal-field $\mu$SR relaxation measurements at low fields reveal a wide distribution of muon relaxation rates and divergences in the frequency dependence of spin correlation functions in the NFL systems UCu$_{5-x}$Pd$_x$ and CePtSi$_{1-x}$Ge$_x$. These divergences seem to be due to slow dynamics associated with quantum spin-glass behavior, rather than quantum criticality as in a uniform system, for two reasons: the observed strong inhomogeneity in the muon relaxation rate, and the strong and frequency-dependent low-frequency fluctuation observed in $\mathrm{U(Cu,Pd)_5}$ and $\mathrm{CePt(Si,Ge)}$. In the NFL materials~CeCu$_{5.9}$Au$_{0.1}$, Ce(Ru$_{0.5}$Rh$_{0.5}$)$_2$Si$_2$, CeNi$_2$Ge$_2$, and YbRh$_2$Si$_2$ the low-frequency weight of the spin fluctuation spectrum is much weaker than in the disordered NFL systems. 
\end{abstract}
\begin{keyword}
Non-Fermi liquids, disorder-driven mechanisms, quantum critical points, CePtSi$_{1-x}$Ge$_x$
\end{keyword}
\end{frontmatter}

The observed breakdown of Landau's Fermi liquid paradigm in a number of metallic systems, among them certain f-electron heavy-fermion materials, has led to an explosion of effort to understand this non-Fermi liquid (NFL) behavior \cite{ITP96els,Stew01,VNvS01}. From the outset a number of mechanisms have been proposed to explain NFL phenomena; most work has concentrated on effects of a quantum critical point (QCP) separating magnetically ordered and paramagnetic phases at zero temperature. Early theories ignored the fact that most heavy-fermion alloys exhibiting such phenomena (Y$_{1-x}$U$_x$Pd$_3$, UCu$_{5-x}$Pd$_x$, \dots) were disordered by chemical substitution. 

After NMR linewidth measurements in UCu$_4$Pd and UCu$_{3.5}$Pd$_{1.5}$ demonstrated that the magnetic susceptibility in these materials was strongly inhomogeneous \cite{BMLA95}, two broad classes of theories began to address the role of disorder in NFL behavior. In the ``Kondo-disorder'' approach \cite{BMLA95,MDK97} interactions between f moments are ignored and structural disorder gives rise to a broad distribution of Kondo temperatures~$T_K$; NFL behavior arises from low-$T_K$ spins which are not in the Fermi-liquid state. In the ``Griffiths-phase'' picture \cite{CNJ00} spin-spin interactions freeze low-$T_K$ f moments into clusters with a wide distribution of sizes; the larger clusters dominate the susceptibility and lead to divergent behavior as the temperature is lowered. Both Kondo-disorder and Griffiths-phase scenarios seem to be compatible with observed $\mu$SR and NMR linewidths \cite{LMCN00,YMRI02}. 

Recent $\mu$SR spin-lattice relaxation experiments in UCu$_{5-x}$Pd$_x$ \cite{MBHN01,MHBN02} indicated, however, that the U-ion spin dynamics are better described by a cooperative picture in which critical slowing down occurs throughout the sample, rather than at (rare) low-$T_K$ spins or large clusters. The $\mu$SR relaxation rates are nevertheless widely distributed, and the dynamic behavior closely resembles that of spin glasses \cite{KMCL96,KBCL01}. Thus an important question is whether the appropriate model is one in which disorder in the interactions dominates the dynamics, as in a spin glass above the glass temperature, or if, instead, the critical slowing down is still controlled by a QCP. Theoretical treatments exist \cite{GrRo99,GPS00} which combine elements of both these viewpoints and describe a ``quantum spin glass'' with a suppressed or possibly zero glass temperature.

In the meantime NFL systems have been discovered in which disorder does not appear to play an essential role; at ambient pressure these include CeCu$_{5.9}$Au$_{0.1}$ \cite{LPPS94,Loeh96}, Ce(Ru$_{1-x}$Rh$_{x}$)$_2$Si$_2$, $x \approx 0.5$ \cite{YMTK99,YMTO99} CeNi$_2$Ge$_2$ \cite{JPGM96,SBGL96}, and YbRh$_2$Si$_2$ \cite{TGML00}. Thus we can compare spin dynamics in ``disordered'' and ``ordered'' materials. The result of this comparison is evidence that the divergences are due to slow dynamics associated with quantum spin-glass behavior, rather than quantum criticality as previously believed. This evidence consists of (a)~the observation of a wide inhomogeneous distribution of relaxation rates characteristic of disordered systems when the relaxation rate is strong; (b)~the fact that in the ordered NFL compounds~CeNi$_2$Ge$_2$ and YbRh$_2$Si$_2$, and even in CeCu$_{5.9}$Au$_{0.1}$ \cite{AFGS95} and Ce(Ru$_{0.5}$Rh$_{0.5}$)$_2$Si$_2$ \cite{YMTK99}, which are doped alloys, the low-frequency weight of the spin fluctuation spectrum is more than an order of magnitude smaller than in the disordered NFL systems, and (c)~the fact that the residual resistivity of NFL systems is strongly correlated with the rapidity and the inhomogeneity of the muon relaxation. The present paper reviews the data and analyses that support this conclusion.

CePtSi$_{1-x}$Ge$_x$ is a NFL system \cite{SGGO94} with similarities to CeCu$_{6-x}$Au$_{x}$. The end compound CePtSi is a heavy-fermion paramagnet. Ge doping expands the lattice and favors antiferromagnetism. A nonzero N\'eel temperature~$T_N$ appears for $x \gtrsim 0.1$, which is therefore a candidate for a QCP. But magnetic susceptibility and $^{29}$Si NMR studies of CePtSi and CePtSi$_{0.9}$Ge$_{0.1}$ \cite{YMRI02} show that both alloys exhibit strong disorder in the susceptibility above 2 K, which when extrapolated to lower temperatures accounts for much op the NFL specific heat. 

\begin{figure}[ht]
\begin{center}
\epsfig{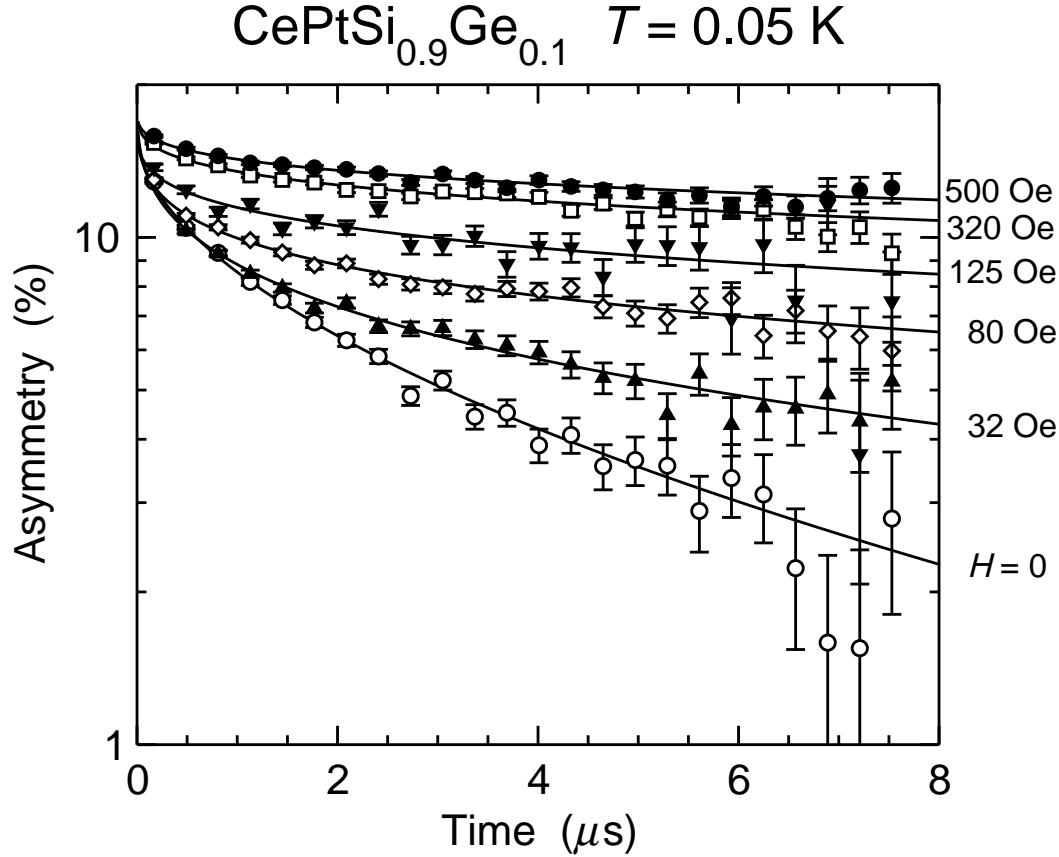}
\end{center}
\caption{Semi-log plots of  LF-$\mu$SR asymmetry data in CePtSi$_{0.9}$Ge$_{0.1}$, $T = 0.05$ K. Curves: fits to stretched exponential $A\exp[-(\Lambda t)^K]$.}
\label{fig:NFLfig1}
\end{figure}

Longitudinal-field $\mu$SR (LF-$\mu$SR) experiments below 1 K in CePtSi$_{1-x}$Ge$_x$, $x = 0$ and 0.1, that exhibit rapid relaxation and {\em time-field scaling\/} \cite{KMCL96,KBCL01} behavior qualitatively similar to that seen in UCu$_{5-x}$Pd$_x$, $x = 1.0$ and 1.5 \cite{MBHN01,MHBN02}. Figure~\ref{fig:NFLfig1} gives the muon decay asymmetry relaxation function for CePtSi$_{0.9}$Ge$_{0.1}$ at $T = 0.05$ K and several field values. 
The data are well described by the stretched exponential function~$A\exp[-(\Lambda t)^K]$, where $A$ is the initial muon decay asymmetry, $\Lambda$ is a characteristic relaxation rate [$1/\Lambda$ is the time at which $G(t)$ decays to $1/\mathrm{e}$ of its initial value], and the exponent~$K < 1$ describes the degree of ``stretching.'' Independently of this fit, the upward curvature of the asymmetry decay data vs time on the semi-log plot of Fig.~\ref{fig:NFLfig1} indicates a distribution of local relaxation rates \cite{KMCL96} and is another clear manifestation of disorder.

\begin{figure}[ht]
\begin{center}
\epsfig{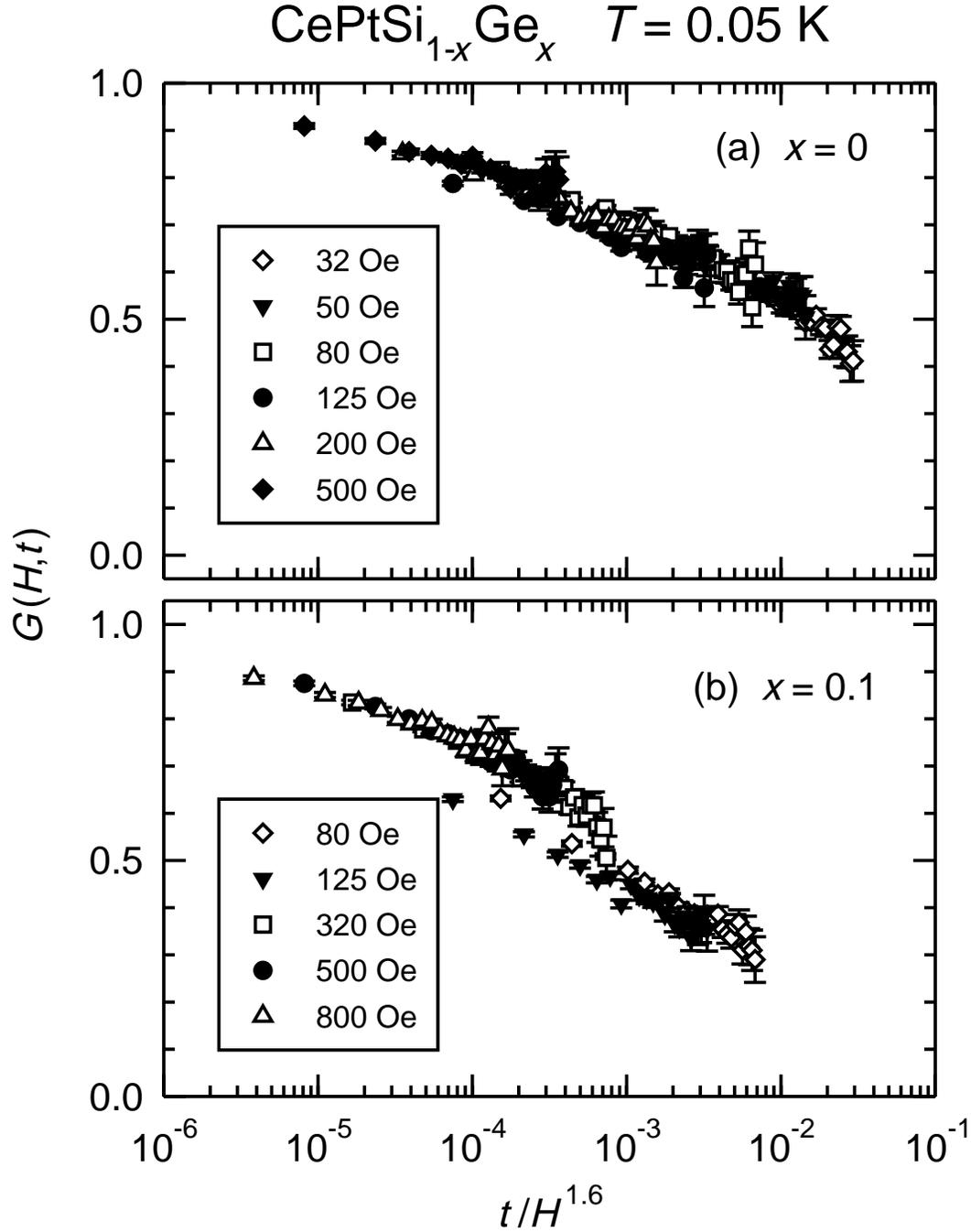}
\end{center}
\caption{Dependence of LF-$\mu$SR relaxation function~$G(H,t)$ on scaling variable~$t/H^{1.6}$ in CePtSi, $T = 0.05$ K.}
\label{fig:NFLfig2}
\end{figure}

Normalized relaxation functions~$G(H,t)$ are plotted in Fig.~\ref{fig:NFLfig2} against the scaling variable~$t/H^{y}$ for CePtSi$_{1-x}$Ge$_x$ at $T = 0.05$ K. 
This procedure is motivated by the finding \cite{MBHN01} that in UCu$_{5-x}$Pd$_x$ the field dependence reflects the dependence of the (inhomogeneous) relaxation rates on muon Zeeman frequency~$\omega_\mu = \gamma_\mu H$. This in turn is related to the frequency dependence of the dynamic susceptibility~$\chi''(\omega, T)$, for which a scaling form was found from neutron scattering studies \cite{AORL95b}. For CePtSi the best scaling [i.e., the most nearly universal relation~$G(H,t) = G(H/T^y)$] is achieved with $y = 1.6 \pm 0.1$. Such a procedure does not require knowledge of the functional form of $G(H,t)$. For CePtSi$_{0.9}$Ge$_{0.1}$ scaling is not obeyed for $H \lesssim 300$ Oe, but above this field scaling is found with the same exponent. The value of this scaling exponent is considerably larger than in UCu$_{5-x}$Pd$_x$ $T \lesssim 1$ K, where $y(x{=}1.0) = 0.35$ and $y(x{=}1.5) = 0.5$--0.7 \cite{MBHN01,MHBN02}.

\begin{figure}[ht]
\begin{center}
\epsfig{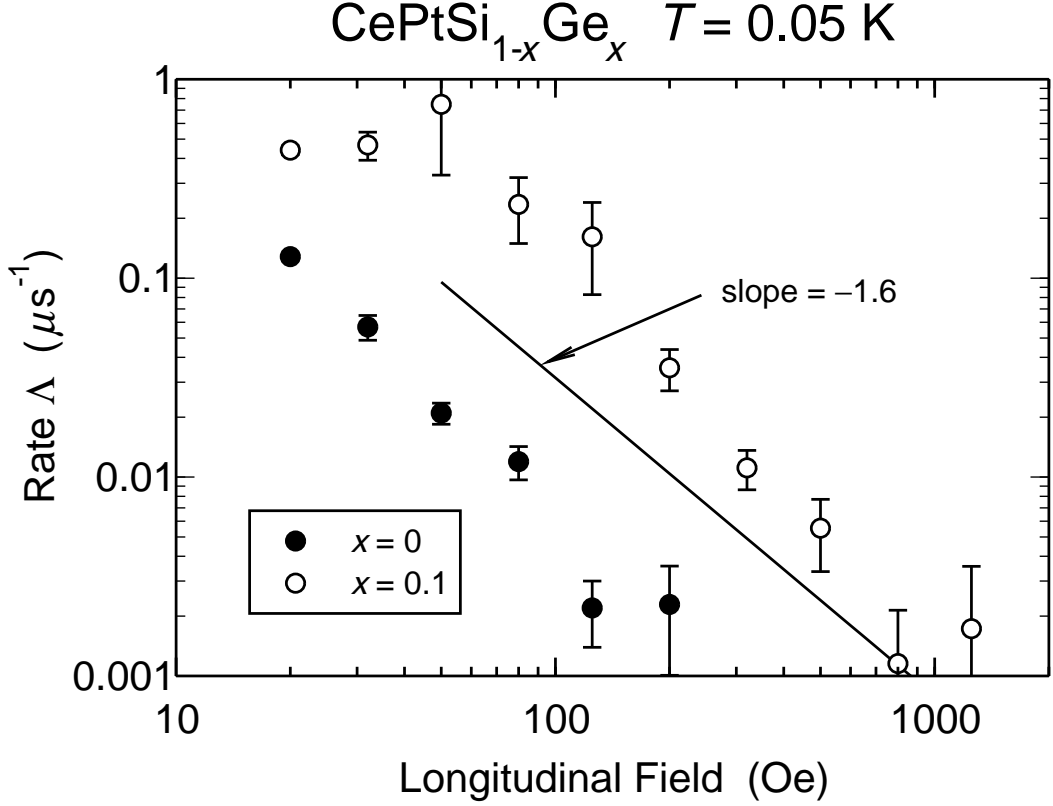}
\end{center}
\caption{Dependence of LF-$\mu$SR stretched-exponential relaxation rate~$\Lambda$ on longitudinal field in the NFL alloys CePtSi$_{1-x}$Ge$_x$, $x = 0$ and 0.1.}
\label{fig:NFLfig3}
\end{figure}

The stretched-exponential fits of Fig.~\ref{fig:NFLfig1} clarify the situation in CePtSi$_{1-x}$Ge$_x$. The field dependence of the relaxation rate~$\Lambda$ is plotted in Fig.~\ref{fig:NFLfig3} for $x = 0$ and 0.1. For both concentrations $K$ is constant at about 0.25 for fields above $\sim 50$~Oe. For $x = 0$ $\Lambda$ varies as $H^{-y}$, $y \approx 1.6$, whereas for $x = 0.1$ this is true (with roughly the same exponent) only above a crossover field of about 100~Oe.
This behavior and the fact that $y$ is greater than 1 are both consistent with a stretched-exponential spin autocorrelation function~$q(t) = \langle S(t)S(0)\rangle = \exp[-(t/\tau_c)^{y-1}]$ \cite{KMCL96,KBCL01}, rather than the power-law~$q(t) = t^{1-y}$ ($y < 1$) found in UCu$_{5-x}$Pd$_x$. The crossover from $\Lambda \approx \rm const.$ to $\Lambda \propto H^{-y}$ occurs for $\omega_\mu\tau_c \sim 1$. The data of Fig.~\ref{fig:NFLfig3} indicate that $\tau_c$ is considerably longer for CePtSi than for CePtSi$_{0.9}$Ge$_{0.1}$, due perhaps to less pinning of fluctuations in the more ordered end compound. The origins of these different behaviors are related to details of the microscopic mechanism for the disordered spin dynamics, and are not well understood.

\begin{figure}[ht]
\begin{center}
\epsfig{file=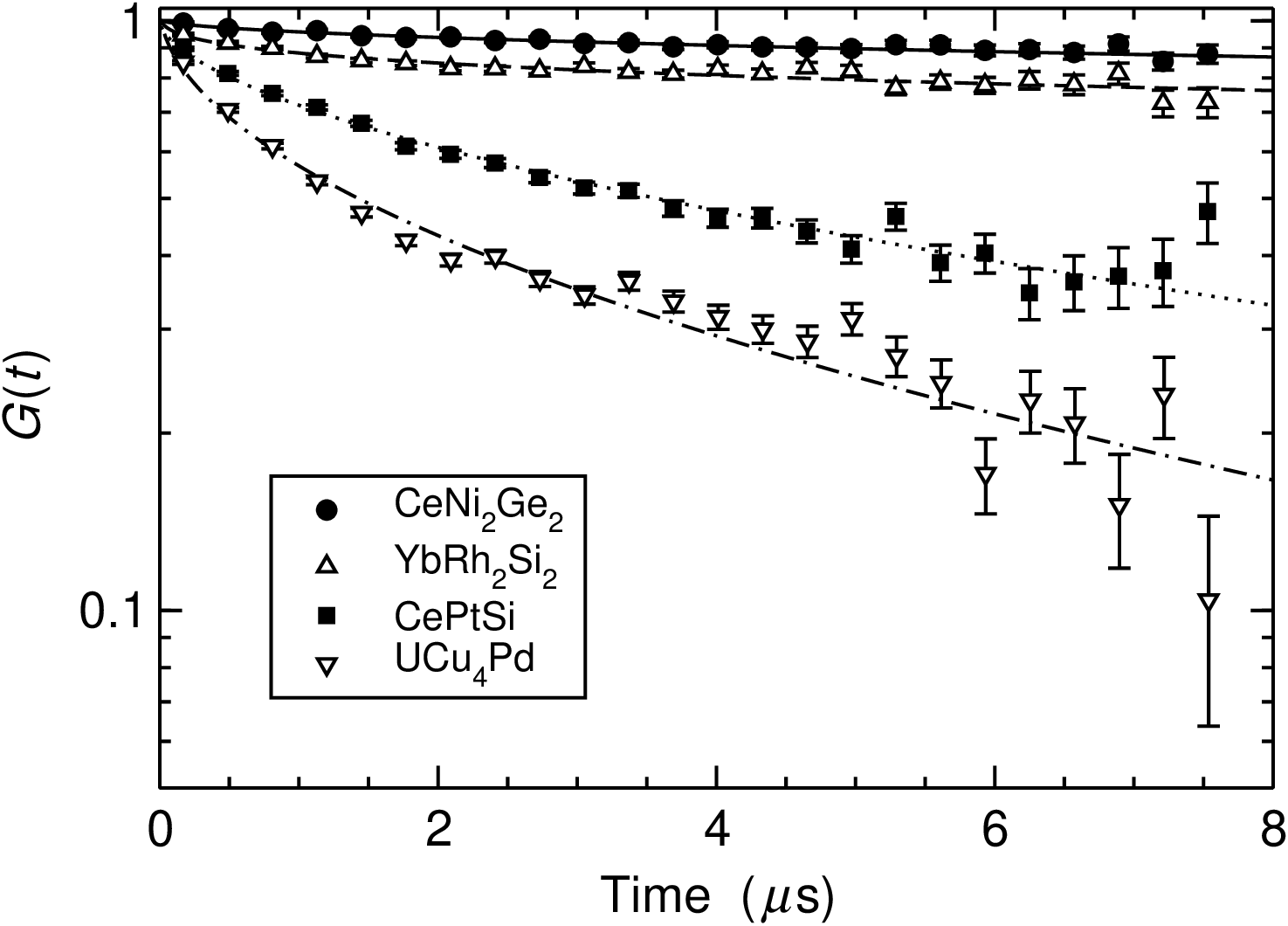,width=\columnwidth}
\end{center}
\caption{LF-$\mu$SR relaxation functions~$G(t)$ in NFL materials. ({\large$\bullet$})~CeNi$_2$Ge$_2$, $T = 0.02$ K, $H = 0$. ({\small$\triangle$})~YbRh$_2$Si$_2$, $T = 0.02$ K, $H = 32$ Oe. ($\scriptstyle \blacksquare$)~CePtSi, $T = 0.05$ K, $H = 32$ Oe. ($\triangledown$)~UCu$_4$Pd: $T = 0.05$ K, $H = 20$ Oe.}
\label{fig:NFLfig4}
\end{figure}

The question then arises as to whether the time-field scaling is due to disordered glassy spin dynamics or to a QCP. We first look at LF-$\mu$SR relaxation in a number of NFL systems in Figure~\ref{fig:NFLfig4}.
If necessary a small longitudinal magnetic field has been applied to ``decouple'' any nuclear dipolar field so that the relaxation is purely dynamic. The estimated $G(t)$ from relaxation rate data for CeCu$_{5.9}$Au$_{0.1}$ \cite{AFGS95} and Ce(Ru$_{0.5}$Rh$_{0.5}$)$_2$Si$_2$ \cite{YMTK99} are comparable to that for CeNi$_2$Ge$_2$, i.e., the relaxation rates are comparable. CePtSi and UCu$_4$Pd show the most rapid and most nonexponential relaxation, whereas muons in CeCu$_{5.9}$Au$_{0.1}$ \cite{AFGS95}, Ce(Ru$_{0.5}$Rh$_{0.5}$)$_2$Si$_2$ \cite{YMTK99,YMTO99}, CeNi$_2$Ge$_2$, and YbRh$_2$Si$_2$ \cite{IMOK02,IMBH02}, all relax much more slowly. YbRh$_2$Si$_2$ exhibits weak exponential muon relaxation at low temperatures and fields, with a rate~$1/T_1$ that varies inversely with longitudinal field \cite{IMBH02}.

Thus there seem to be two classes of relaxation behavior in NFL materials: UCu$_{5-x}$Pd$_x$ and CePtSi$_{1-x}$Ge$_x$ exhibit rapid and inhomogeneous relaxation, whereas the relaxation rates for the others are much smaller. Is this simply due to differences in characteristic fluctuation rates, or is disorder intimately involved?

By definition NFL metals at low temperatures exhibit no characteristic energy other than the temperature. Nevertheless we can use quantities such as the low-temperature Sommerfeld specific-heat coefficient~$\gamma(T)$ as rough estimates of expected fluctuation rates, and hence of muon relaxation behavior if slow fluctuations are correlated with large values of $\gamma(T)$. Table~\ref{tab:estrate} gives $\gamma(T{=}1\ \mathrm{K})$ for a number of NFL systems, in order from maximum to minimum $\gamma(T{=}1~\mathrm{K})$. 
\begin{table}[b]
\caption{Sommerfeld specific-heat coefficient~$\gamma(T)$ at $T = 1$ K, characteristic temperature~$\pi^2k_B/2\gamma(1$~K), and residual resistivity~$\rho(0)$ for selected NFL compounds.}
\begin{center} \begin{tabular}{lccc} \hline
material & $\gamma(1$~K) & $\pi^2k_B/2\gamma(1$~K) & $\rho(0)$ \\
& (J mol$^{-1}$ K$^{-2}$) & (K) & \ $(\mu\Omega$-cm) \\ \hline
CeCu$_{5.9}$Au$_{0.1}$ & 1.0 \protect\cite{LPPS94}& 8 & 30--60 \protect\cite{Loeh95} \\
CePtSi$_{0.9}$Ge$_{0.1}$ \protect\cite{HMKS92} & 1.0 & 8 & $--$ \\
CePtSi & 0.75 \protect\cite{LeSh87,KSKG90} & 12 & 100 \protect\cite{OFTS92} \\
Ce(Ru$_{0.5}$Rh$_{0.5}$)$_2$Si$_2$ & 0.5 \protect\cite{LCBE87} & 16 & 65 \protect\cite{TGOT01} \\
YbRh$_2$Si$_2$ & 0.5 \protect\cite{TGML00} & 16 & 2.4 \protect\cite{TGML00} \\
UCu$_4$Pd & 0.45 \protect\cite{AnSt93} & 18 & 280 \protect\cite{CHMA97} \\
UCu$_{3.5}$Pd$_{1.5}$ & 0.38 \protect\cite{AnSt93} & 25 & 180 \protect\cite{CHMA97} \\
CeNi$_2$Ge$_2$ & 0.32 \protect\cite{KLCG88} & 25 & 0.4 \protect\cite{SSTL00} \\ \hline
\end{tabular} \end{center}
\label{tab:estrate}
\end{table}
The corresponding characteristic temperature~$\pi^2k_B/2\gamma(1$~K), i.e., the Fermi temperature in a free-electron gas, is also given. This is admittedly only a crude gauge of spin fluctuation rates in these systems. Nevertheless, in a conventional picture higher values of $\pi^2k_B/2\gamma(1$~K) would imply faster spin fluctuations, more motional narrowing, and slower muon relaxation rates. The residual resistivity~$\rho(0)$ is also included in Table~\ref{tab:estrate} as a measure of the effect of disorder, intrinsic or extrinsic, on the electronic structure.

Comparison of Table~\ref{tab:estrate} and Fig.~\ref{fig:NFLfig4} shows little correlation between muon relaxation behavior and $\gamma(1$~K). Both CeCu$_{5.9}$Au$_{0.1}$ and Ce(Ru$_{0.5}$Rh$_{0.5}$)$_2$Si$_2$, with relatively large values of $\gamma(1$~K) but with little effect of disorder, relax muons very slowly, whereas muons are relaxed rapidly in UCu$_4$Pd, which has a small value of $\gamma(1$~K) but a large residual resistivity. In contrast, there is an excellent correlation between the residual resistivity (Table~\ref{tab:estrate}) and the strength (and inhomogeneity) of the muon relaxation (Fig.~\ref{fig:NFLfig4}). 

We therefore conclude that rapid relaxation in NFL materials is due to disorder-induced increase of low-frequency spectral weight of the spin fluctuations. Quantum critical behavior by itself seems not to generate strong low-frequency spin fluctuations. This behavior must be taken into account in further attempts to understand NFL behavior in disordered f-electron metals.

We wish to thank A. Amato, D. Arsenau, C. Baines, M. Good, D. Herlach, B. Hitti, and S. Kreitzman for help with these experiments, and to M. Aronson, A. Castro Neto, F. Steglich, L. Pryadko, and Q. Si for useful discussions. This work was supported in part by the U.S. NSF, Grant nos.~DMR-0102293 (UC Riverside) and DMR-9820631 (CSU Los Angeles), by the COE Research Grant-in-Aid (10CE2004) for Scientific Research from the Ministry of Education, Sport, Science, and Technology of Japan, and by the Netherlands NWO and FOM, and was performed in part under the auspices of the US DOE.

\end{document}